\documentclass[twocolumn,showpacs]{revtex4}

\usepackage{color}
\usepackage{graphicx}
\usepackage{dcolumn}
\usepackage{amsmath}

\begin{document}
\bibliographystyle{revtex}


\title[Short Title]{Topological characterization of neutron star crusts}

\author{C. O. Dorso and P. A. Gim\'enez Molinelli}

\affiliation{Departamento de F\'isica, FCEN, Universidad de Buenos
Aires, N\'u\~nez, Argentina}

\author{J. A. L\'opez}

\affiliation{Department of Physics, University of Texas at El
Paso, El Paso, Texas 79968, U.S.A.}

\date{\today}

\begin{abstract}
Neutron star crusts are studied using a classical molecular
dynamics model developed for heavy ion reactions. After the model
is shown to produce a plethora of the so-called ``pasta'' shapes,
a series of techniques borrowed from nuclear physics, condensed
matter physics and topology are used to craft a method that can be
used to characterize the shape of the pasta structures in an
unequivocal way.
\end{abstract}

\maketitle

\section{Introduction}\label{intro}

Neutron stars created in the death of a massive star are composed
of a dense core containing an excess of neutrons over protons,
thus justifying the name.  With a mass between 1 and 3 solar
masses and a radius of about 10 km, the stars are topped with a
crust of about a kilometer thick where the $\beta$
decayed-produced neutrons form neutron-rich nuclear matter
immersed in a sea of electrons. The crust density ranges from
normal nuclear density ($\sim 3\times10^{14}$ $g/cm^3$) at a depth of about $1 \ km$, to
the neutron drip density ($\sim 4\times10^{11}$ $g/cm^3$) at about
$ 1/2 \ km$, to a light mix of neutron-rich nuclei with densities
decreasing practically to zero in the neutron star envelope.
Likewise the proton-to-nucleon ratio also varies from $\sim 0.25$
to $\sim 0.5$ through the crust~\cite{haensel}, and the
temperature from cold nuclear matter to about $1 \ MeV$. The study
of the structure of such crust, is the purpose of the present
work.

Studies of low density nuclear matter have found that the
attractive-repulsive interplay of nuclear and Coulomb forces drive
low-density nuclear matter to form non-uniform structures which
are collectively known as ``nuclear pasta''. Such arrangements go
from condensed phases with voids filled with nuclear gas, to
``lasagna-like'' layers of nuclei surrounded by gas, to
``spaghetti-like'' rods embedded in a nuclear gas, to ever
decreasing ``meatball-like'' clumps which practically dissolve
into a gaseous phase~\cite{4}.

Early investigations have used static models which rely mostly on
energy considerations to determine the structures that are most
energetically favorable.  Among the various theories proposed, the
ones used most recurrently are the Compressible Liquid Drop
Model~\cite{L3,L54,L30}, the Extended Thomas-Fermi
Model~\cite{4,L79,L16,Avancini}, and the Hartree-Fock
method~\cite{magierski,L76,L72,L30,L62,Newton}.

On the other hand, there are dynamical models that go beyond mean
fields to predict the formation of the pasta phases as an
asymptotic equilibrium state resulting from an evolution of a
dynamical system. The most used methods are the semiclassical
molecular dynamics~\cite{horo35801,horo_lambda, horo65806} and the
quantum molecular
dynamics~\cite{7,Oyamatsu1984,Watanabe2009,Nakazato,gw-2002,9}.

On the general composition of the pasta, most models agree on the
formation of varying structures at subnormal densities but not on
how the sequence the phases arises. Since the physical mechanism
responsible for the phase transition pattern is a subtle interplay
between Coulomb and nuclear energies which varies only a few
$keV/fm^{3}$ between phases, the precise transition pattern is
easily altered by the ingredients of the theoretical models. The
fact that the Coulomb interaction between the electron sea and the
nucleons --whose screening effect stabilizes the overall system
modifying its structure-- must be treated under different
approximations in different models~\cite{9} complicates any
cross-model comparison even more.

An additional problem which makes comparisons between models
difficult or impossible in some cases, is the lack of a
quantifiable characterization of different pasta phases. The
identification of phases has been done mainly through visual
inspections of snapshots of spatial nucleon distributions obtained
from calculations~\cite{Wata-2010}. The QMD approach of
reference~\cite{7}, for example, produces nuclear holes, slabs,
cylinders and spheres similar to those predicted by the Thomas
Fermi model~\cite{4,L79,L16}, but at different densities and
temperatures.

In spite of this, the pasta phases have been characterized
globally.  For instance, static models have been used to calculate
average densities~\cite{Avancini} and volume fractions of the
different phases~\cite{Maruyama-2005}. Pasta bulk properties, such
as the shear viscosity~\cite{Horo-2008} and diffusion
coefficients~\cite{menezes} have also been obtained using
molecular dynamics simulations. Refined studies have used radial
correlation functions to characterize the nucleon
distributions~\cite{Watanabe2009} and the pasta structure factor
to study charge density fluctuations~\cite{horo-2006}.

More recently, shape characterizations were attempted both in
dynamical simulations and with static models; the former use
topological measures such as the Minkowski functionals and the
Euler characteristics~\cite{Wata-2010}, while the latter modified
the liquid drop model with a curvature correction to detect
structure shape changes~\cite{nakazato-2011}.

Thus the motivation of the present study: how to achieve a precise
enough characterization of the pasta phases? What property can be
used to signal a change of pasta phase? The purpose of the present
work is to construct the instruments needed to properly quantify
the pasta structures.

Taking advantage of the microscopic details produced by a
classical molecular dynamics model, this investigation combines
the power of cluster detection algorithms used in nuclear
collisions with indicators borrowed from condensed matter physics
and topology to detect the transitions between pasta phase
structures in a quantitative way.

After an introduction of the model, we will introduce a series of
techniques used to classify the pasta structures that will help us
reach our goal. Starting from global measures to understand the
cluster composition (fragment size distribution, nucleon mobility
and persistence, fragment isotopic composition and radial
distribution function), we will progress into topological tools
(Minkowski functionals) that will allow us to characterize the
shape of the pasta structures as well as to detect changes between
them. A final discussion of the results will then help us reach a
series of conclusions and to draw an outlook of the future tasks.

\section{Nucleon dynamics}\label{nd}
To study the  structure of stellar crusts is necessary understand
the behavior of nucleons at the proper densities, temperatures and
proton-to-neutron ratios; such knowledge comes from the study of
heavy ion fragmentations. The initial statistical studies of
nuclear collisions of the 1980's~\cite{Bon96,randruplopez},
rapidly gave way to dynamical theories based on classical,
semiclassical and quantum approaches.

The semiclassical models use the Boltzmann-Uehling-Uhlenbeck
equations~\cite{Ueh33} to track the time evolution in phase space
of the probability of finding a particle moving in a mean field.
On the other hand, the quantum molecular dynamics models ($QMD$)
solve the equations of motion of nucleon wavepackets moving within
mean fields. Unfortunately, these theories either do not lead to
cluster formation or yield a poor description of cluster
properties and both must resort to the use of all sorts of
extraneous techniques such as adding fragments by hand, coupling
to ``afterburners'' to produce secondary decays~\cite{Pol05}, and
introducing hidden adjustable parameters (e.g. width of
wavepackets, number of test particles, modifications of mean
fields, effective masses and cross sections, etc.) to satisfy the
operator's taste.

These problems are either non-existent or much reduced in
classical models.  Classical dynamical models generally solve
Newton's equations of motion to track individual nucleons moving
under two-body potentials; coupled to cluster recognition
algorithms these calculations yield microscopic views of nuclear
reactions as well as of nuclear structures. The only apparent
disadvantage of the classical models would be the lack of quantum
effects, such as the Pauli blocking; fortunately, in stellar
environments the very small nucleon energies lead to frozen-like
structures where the blocking of momentum-transferring collisions
ceases to be relevant. [See Ref.~\cite{horotesis} for a
calculation of the mean thermal wavelength of a nucleus in stellar
conditions to justify the use of a classical approach.]

Let this rather long preamble serve to justify extending the use
of a classical model designed for nuclear reactions to the study
neutron star crusts. In this work we use a classical model to
obtain a detailed microscopic picture of the pasta structures and
be able to detect transitions between phases.

\subsection{Classical Molecular Dynamics}\label{md}

We use a molecular dynamics code combined with algorithms
for cluster recognition. Our classical molecular
dynamics model, $CMD$~\cite{14a}, is based on the pioneering work
of Pandharipande~\cite{pandha} and has been very fruitful in
nuclear studies of, among other phenomena, neck fragmentation
\cite{Che02}, phase transitions \cite{16a,Bar07}, critical
phenomena~\cite{CritExp-1,CritExp-2}, the caloric
curve~\cite{TCalCur,EntropyCalCur}, and isoscaling \cite{8a,Dor11}
all without any adjustable parameters. Readers are directed to
these references for further details on the model; here only a
brief synopsis will be presented along with its extension to
infinite systems.

In a nutshell, $CMD$ treats nucleons as classical particles
interacting through a two-body potential and solves the coupled
equations of motion of the many-body system to obtain the time
evolution of all particles.  Since the $(\mathbf{r},\mathbf{p})$
information is known for all particles at all times, it is
possible to know the structure of the nuclear medium from a
microscopic point of view.

$CMD$ uses the phenomenological potentials developed by
Pandharipande~\cite{pandha}:
\begin{eqnarray*}
V_{np}(r) &=&V_{r}\left[ exp(-\mu _{r}r)/{r}-exp(-\mu
_{r}r_{c})/{r_{c}}
\right] \\
& &\ \mbox{}-V_{a}\left[ exp(-\mu _{a}r)/{r}-exp(-\mu
_{a}r_{a})/{r_{a}}
\right] \\
V_{NN}(r)&=&V_{0}\left[ exp(-\mu _{0}r)/{r}-exp(-\mu _{0}r_{c})/{
r_{c}}\right] \ , \label{2BP}
\end{eqnarray*}
where $V_{np}$ is the potential between a neutron and a proton and
it is attractive at large distances and repulsive at small ones,
and $V_{NN}$ is the interaction between identical nucleons and it
is purely repulsive.  Notice that no bound state of identical
nucleons can exist, also notice that, at a difference from
potentials used by other models~\cite{horo_lambda}, these
potentials have a hard core.

The cutoff radius is $r_c=5.4$ $fm$ after which the potentials are
set to zero. Two sets of values for the Yukawa parameters $\mu_r$,
$\mu_a$ and $\mu_0$ were fixed by Pandariphande to correspond to
infinite-nuclear matter systems with an equilibrium density of
$\rho_0=0.16 \ fm^{-3}$, a binding energy $E(\rho_0)=16$
MeV/nucleon and compressibility of about $250 \ MeV$ (``Medium'') or
$535 \ MeV$ (``Stiff'')~\cite{pandha}. In the past, a combination of
Monte Carlo and molecular dynamics techniques was applied within a
statistical formalism to obtain neutron start crust
properties~\cite{horo_lambda}.

\subsection{Simulating the neutron star crust}\label{cp}

To study the neutron start crust we use $CMD$ to simulate an
infinite medium. Systems with $2000$ or $3000$ nucleons were
constructed and replicated with periodic boundary conditions in 26
surrounding cells.  In particular, the proton ratios used were of
$x=Z/A=0.5$ ($1000$ neutrons and $1000$ protons) or $0.3$ ($2000$
neutrons and $1000$ protons). The cubical box size used was
adjusted as to achieve densities between $\rho=0.01 \ fm^{-3}$ (about
$\rho_0/15$) and $\rho_0$.

\begin{figure}[t]  
\begin{center}
\includegraphics[width=3.5in, height=3.5in]{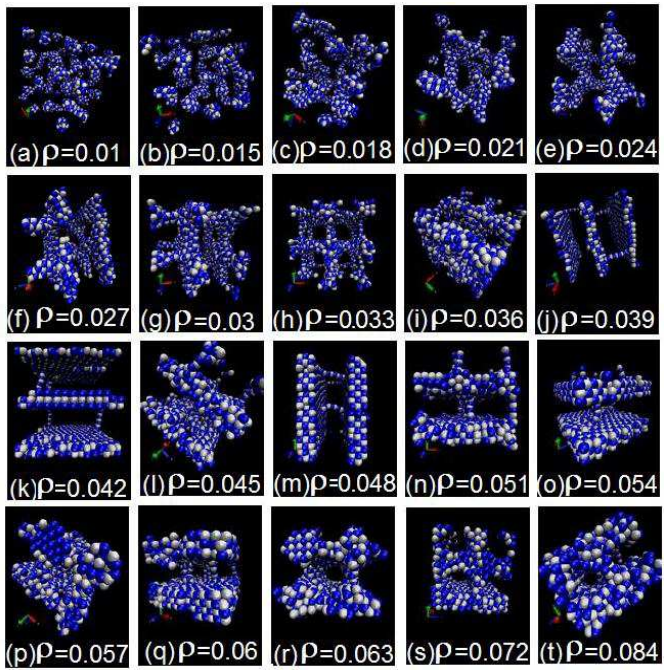}
\end{center}
\caption{(Color online) Sm\"org{\aa}sboard of pasta shapes corresponding to the
densities shown and to $x=0.5$ and $T=0.1 \ MeV$.}\label{Pasta}
\end{figure}

As the crust is expected to be embedded in a degenerate electron
gas produced by weak decays during the supernova explosion, it is
necessary to take into account its Coulomb interaction. Although
the nucleon-electron system is overall neutral and
$\beta$-equilibrated, the infinite Coulomb range requires the use
of some approximation; two common approaches are the Thomas-Fermi
screened Coulomb potential (used in $QMD$ in~\cite{7}) or the
Ewald summation procedure~\cite{wata-2003}. Although $CMD$ is able
to operate under either approximation, in this work the former is
adopted (see~\cite{Dor12} for a comparison of methods under $CMD$).

Approximating the electron gas as a uniform ideal Fermi gas at the
same number density as the protons, its effect can be included in
the nucleon's equations of motion by means of the screened Coulomb
potential obtained from the Poisson equation:
\[V_C^{(Scr)}(r)=\frac{e^2}{r}\exp(-r/\lambda)\]
where the relativistic Thomas-Fermi screening length is:
$\lambda=(\pi^2/2e)\left[k_F^2
\left({k_F^2+m_e^2}\right)\right]^{-\frac{1}{4} }$, $m_e$ is the
electron mass, $k_F=\left( 3\pi^2\rho_e\right)^{1/3}$ is the
electron Fermi momentum, and $\rho_e$ is the electron gas number
density equal to that of the protons. The size of the simulation
cell, $L=\left( A/\rho\right)^{ 1/3 }$, should be significantly
larger than $\lambda$; in our case we satisfy such requirement
using the prescription of~\cite{horo_lambda} and setting
$\lambda=10 \ fm$.

\begin{figure}[t]  
\begin{center}
\includegraphics[width=4.3in]{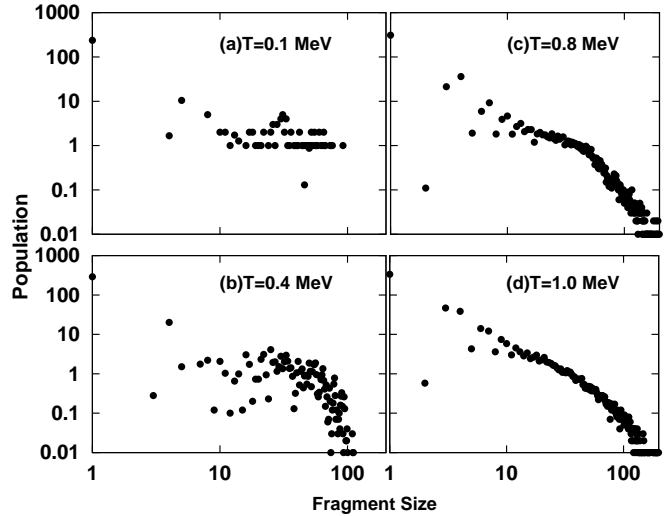}
\end{center}
\caption{Temperature evolution of the fragment size distribution
obtained from $200$ configurations with $x=0.3$
and $\rho\approx\rho_0/10$.}\label{MassDist-T}
\end{figure}

The trajectories of individual nucleons, now governed by the
Pandharipande and the screened Coulomb potentials, are then
tracked using a Verlet algorithm with energy conservation of
$\mathcal{O}$($0.01 \%)$. The system is force-heated or cooled
using isothermal molecular dynamics with the Andersen thermostat
procedure~\cite{andersen} which gradually cools in small
temperature steps while reaching thermal
equilibrium at every step. We focus in the range of $T=0.1$ to
$1.0 \ MeV$; although this last temperature is large for stellar
crusts, in terms of the nucleon dynamics it practically
corresponds to a frozen state.

\section{Characterizing the crust}
At a difference from most $QMD$ simulations, which tend to track
individual evolutions, here we obtain reliable statistics by
sampling 200 times each configuration with specific $x$, $\rho$,
and $T$ conditions. Figure~\ref{Pasta} shows an sm\"org{\aa}sboard
of Italian delicacies produced by $CMD$ with $x=0.5$, $T=0.1 \
MeV$ and twenty different densities; please notice that for
clarity the figures do not show single nucleons, i.e. the gaseous
phase. In spite of their beauty, one cannot use those figures to
properly characterize the pasta shapes, for that one must resort
to other, less visually attractive, techniques.

\begin{figure}[t]  
\begin{center}
\includegraphics[width=4.1in]{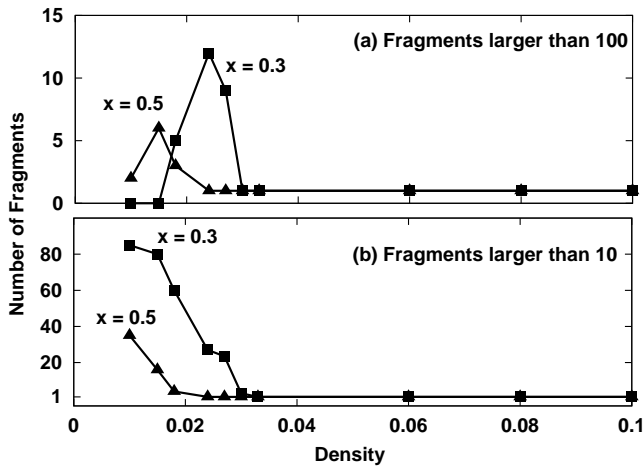}
\end{center}
\caption{Relative abundance of large clusters as a
function of density for $x=0.3$ and $0.5$ at $T=0.1 \
MeV$.}\label{Multiplicity}
\end{figure}

On each of the configurations achieved, the nucleon
$(\mathbf{r},\mathbf{p})$ information is recorded and used later
to identify clusters and to characterize the structure by means of
the liquid structure function and the Minkowski functionals.

\subsection{Cluster composition}\label{clustercomposition} The nucleon positions are used to identify clusters by means of the ``Minimum Spanning Tree''
($MST$) algorithm refined for nucleon dynamics
in~\cite{Dor95,Str97}. In summary, $MST$ looks for correlations in
configuration space: a particle $i$ belongs to a cluster $C$ if
there is another particle $j$ that belongs to $C$ and
$|\mathbf{r}_i-\mathbf{r}_j| \leq r_{cl}$, where $r_{cl}$ is a
clusterization radius which is set to $3.0 \ fm$. In spite of
using only $r$-space correlations, $MST$ yields accurate results
in the case of stellar crusts due to the low temperatures and
small momentum transfer, and thus here it is preferred over other
more robust cluster-detection algorithms (such as the ``Early
Cluster Recognition Algorithm'', $ECRA$~\cite{dor-ran}, which take
into account relative momenta and binding energies). In our
case of periodic boundary conditions, the $MST$ method was
modified to recognize fragments that extend into adjacent cells.

\begin{figure}[t]  
\begin{center}
\includegraphics[width=3.5in]{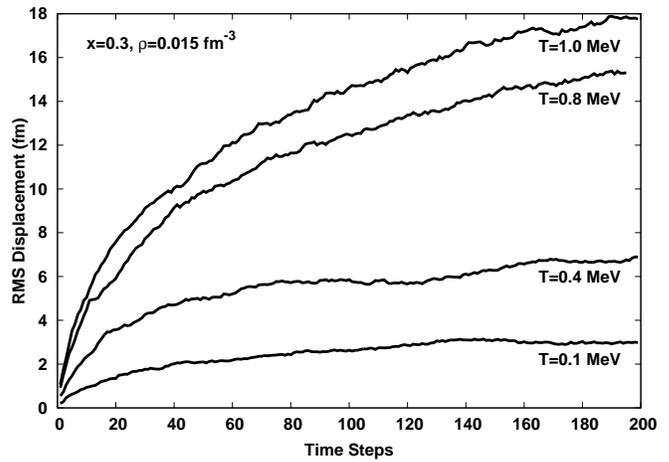}
\end{center}
\caption{Root mean square displacement of the nucleons
from their original position as a function of the simulation time
steps for systems with $\rho=0.015 \ fm^{-3}$ and $x=0.3$ at the listed temperatures.
}\label{Disp}
\end{figure}

Figure~\ref{MassDist-T} shows examples of the fragment population
obtained with $CMD-MST$ at $x=0.3$, $\rho=0.015 \ fm^{-3}$ and
four different temperatures; of particular interest is the
evolution of the clustering as a function of the temperature as it
shows a trend opposite to that observed in nuclear collisions.
The figure shows a typical evolution for the range $0.1 \ MeV < T
< 0.1 MeV$ and, as it can be clearly observed, the large fragment
multiplicity increases with $T$.  This is in opposition to what
happens in heavy ion reactions at high energies where heavier
fragments shrink in size by particle evaporation during the final
expansion stage of the reaction.  In the case of infinite systems,
however, the lack of expansion (and lack of a reduced pressure)
makes evaporation less probable and, combined with the possibility
of connecting fragments to neighboring cells, it favors the growth
of cluster sizes as soon as the nucleons reach enough mobility
with increasing $T$.

The growth of large fragment multiplicity can also be seen as a
function of the density.  Figure~\ref{Multiplicity} shows a
typical behavior of the relative multiplicity of large clusters,
$A>100$ and $A>10$, obtained at different densities for both
$x=0.3$ and $0.5$ at $T=0.1 \ MeV$.  As the density increases, the
number of clusters of $A>100$ increases practically linearly with
$\rho$ up until a single large fragment is formed; smaller
clusters of $A>10$ are abundant at low densities but decrease for
larger densities.  The density at which the number of large
clusters condense into a single one can be thought of as a
``percolation'' density, this value, of course, depends on the
simulation parameters such as number of particles, cell size,
temperature, etc.; for the cases shown, the percolation densities
are $\rho\approx0.03 \ fm^{-3}$ for $x=0.3$ and $\rho\approx 0.024
\ fm^{-3}$ for $x=0.5$ at $T=0.1 \ MeV$.

\begin{figure}[t]  
\begin{center}
\includegraphics[width=3.5in]{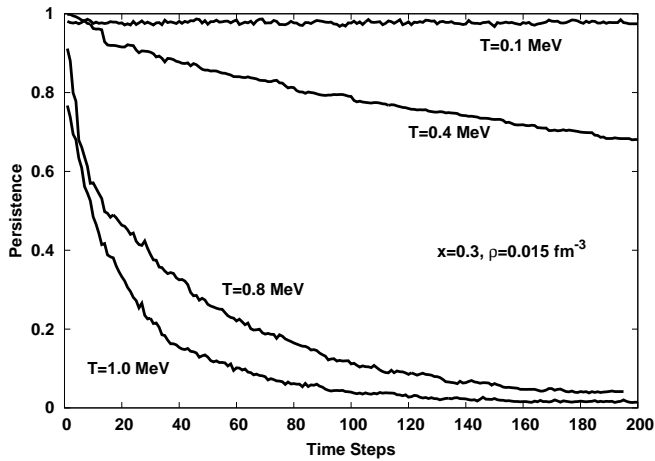}
\end{center}
\caption{Time evolution of the persistency for the listed
temperatures as a function of the simulation time steps for
$x=0.3$ and $\rho=0.015 \ fm^{-3}$. }\label{Persis}
\end{figure}

The dynamics of the nucleons within systems in equilibrium can be
gauged through their average displacement as a function of
``time'', i.e. through the time steps of the simulation.
Figure~\ref{Disp} shows the root mean square displacement of the
nucleons from their original position in $200$ time steps; as a
metric one must remember that the range of the potential is $5.4 \
fm$ and -as we will see in section~\ref{pair} -- the interparticle
distance at these densities is of the order of $1.7 \ fm$.  The
increment in mobility as a function of the temperature is obvious.

Likewise, the microscopic stability of the clusters can be
quantified through the ``persistency''~\cite{DorStr95,Lop00} which
measures the tendency of members of a given cluster to remain in
the same cluster.  Figure~\ref{Persis} shows the time evolution of
the persistency for systems with $\rho=0.015 \ fm^{-3}$ and
$x=0.3$ at the listed temperatures; notice that a persistency of
$\sim 1$ indicates that most of the particles remain in the same
cluster, while smaller values indicate a larger exchange rate.
The anti-correlation between this and the previous figure is
clear, more mobility implies less persistency, and viceversa.

\begin{figure}[t]  
\begin{center}
\includegraphics[width=4.in]{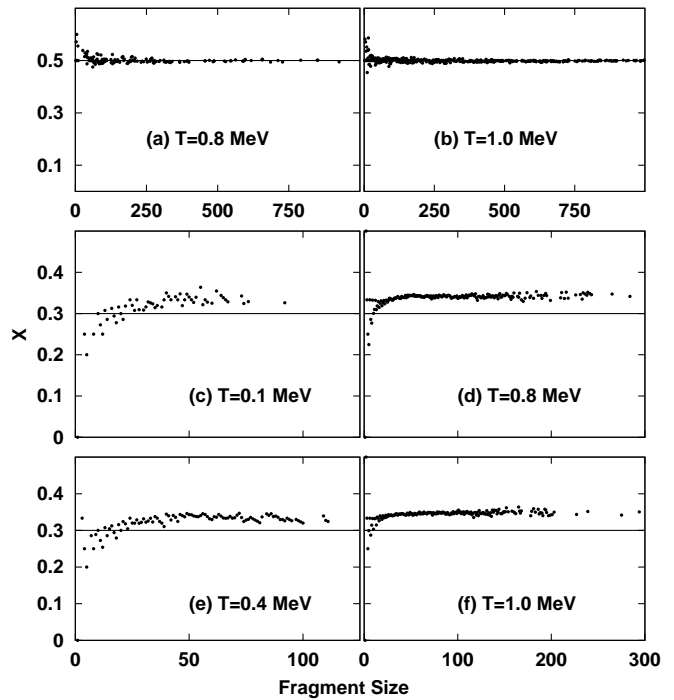}
\end{center}
\caption{Example of the $x$ value ($x=p/A$) of the clusters versus
their size, corresponding to systems with $\rho=0.015 \ fm^{-3}$
and $x=0.5$ (top two plots) and $x=0.3$ (bottom four plots) at the
temperatures listed.}\label{x-of-frags}
\end{figure}

Another interesting descriptor is the isospin content of the
clusters produced.  By keeping track of the number of protons and
neutrons on each fragment it is possible to determine the $x$
value for each cluster.  An example of this is shown in
figure~\ref{x-of-frags} where the $x$ content of the fragments is
plotted as a function of the mass of the clusters obtained at a
density of $\rho=0.015 \ fm^{-3}$ and with $x=0.5$ (top two plots)
and $x=0.3$ (bottom four). Several effects are noticeable: small
clusters ($A\lesssim10$) tend to have less protons than the
average resulting in smaller $x$ values; for the case of $x=0.3$
there is a prominent binding of one proton to two neutrons which
results in clusters of all sizes with values of $x\approx1/3$; the
previous effect is not present in the case of $x=0.5$ in which all
the cluster maintain their $x$ values around $1/2$.

We close this subsection noticing that, in spite of being a good
indicator of the percolating density, the cluster multiplicity is
a poor descriptor of the pasta shapes. Stepping up in complexity,
we now turn to the radial correlation function to probe the pasta
{\it al dente}.

\subsection{Pair correlation function} \label{pair}
Another global characterization
of the structure of nuclear matter is obtained from the pair
correlation function, $g(r)$, which is the ratio of the average
local density to the global density, $g(r) = \rho(r)/\rho_0$; it
gives information about the spatial ordering of the nuclear
medium.

For computing purposes, the pair correlation function $g(r)$ is
taken as the conditional probability density of finding a particle
at $\mathbf{r}_i+\mathbf{r}$ given that there is one particle at
$\mathbf{r}_i$. Formally,
\[
g(r) = \frac{V}{4\pi r^2 N^2} \left\langle \sum_{i\neq j} \delta
\left( r-r_{ij} \right) \right\rangle ,
\]
where $r_{ij}$ is $|\mathbf{r}_i-\mathbf{r}_j|$. For our case,
this was calculated by constructing histograms of the distances
between particles for several configurations obtained with the
same $x$, $\rho$ and $T$ and then averaging them; to include all
particles and their images the range was extended to $r_{ij} \leq
1.5L$.

\begin{figure}[t]  
\begin{center}
\includegraphics[width=3.2in]{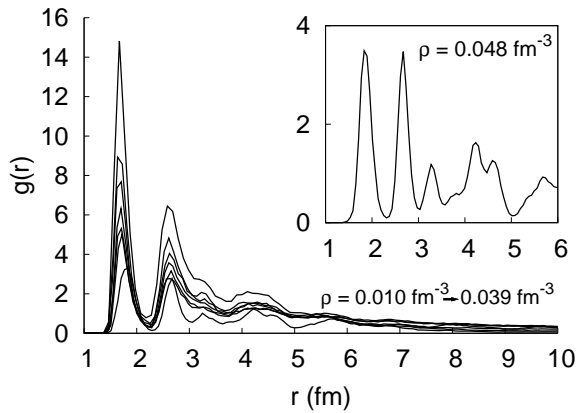}
\end{center}
\caption{Examples of the radial correlation function for $x=0.5$,
$T=0.1 \ MeV$ and various densities. Also shown is the case when
nearest neighbors are as probable as second nearest neighbors
indicating a ``lasagna'' type structure; $r$ is in $fm$.}\label{Radial}
\end{figure}

Figure~\ref{Radial} shows examples of $g(r)$ obtained for cases with
$2,000$ nucleons, $x=0.5$, $T=0.1 \ MeV$ and densities $0.01 <
\rho < 0.039 \ fm^{-3}$.  The inset shows the case $\rho = 0.048 \
fm^{-3}$ when the nearest neighbors are just as probable as second
nearest neighbors signalling the onset of a ``lasagna'' type
structure, cf. Fig.~\ref{Pasta}.

It is worth noticing that, in the case shown, the location of the nearest neighbors
remains constant at $r\sim1.7 \ fm$ at all densities.  This is due
to the fact that --since at subcritical densities the medium is
metastable or unstable, it breaks into a gaseous and a condensed
phase-- the position of the peak of $g(r)$ is an average between
the location of neighbors in the gas-liquid mixture; the condensed
matter at normal density has nearest neighbors at $r\approx1.4 \
fm$~\cite{pandha}.

Once again, in spite of the rich information derived from $g(r)$,
it is still insufficient to tag the phases unequivocally; for this,
other more complex constructs must be borrowed from cosmology and,
ultimately, from topology.

\subsection{Topological constructs}\label{topo}

The most obvious properties of closed surfaces that can be used to
characterize their shapes are the volume $V$, surface area $A$,
and the curvature.  The latter is less trivial than the other two
as it does not hold a unique value for a given shape but varies
from point to point; however, mean curvatures of a closed body can
be obtained by an averaging procedures such as the ``integral mean
curvature'', defined as $H=\int df (R_1+R_2)/2R_1R_2$ where $R_1$
and $R_2$ are the principal radii of curvature of the surface and
$df$ is a differential of area.

\begin{figure}[t]  
\begin{center}$
\begin{array}{cc}
\includegraphics[width=1.6in]{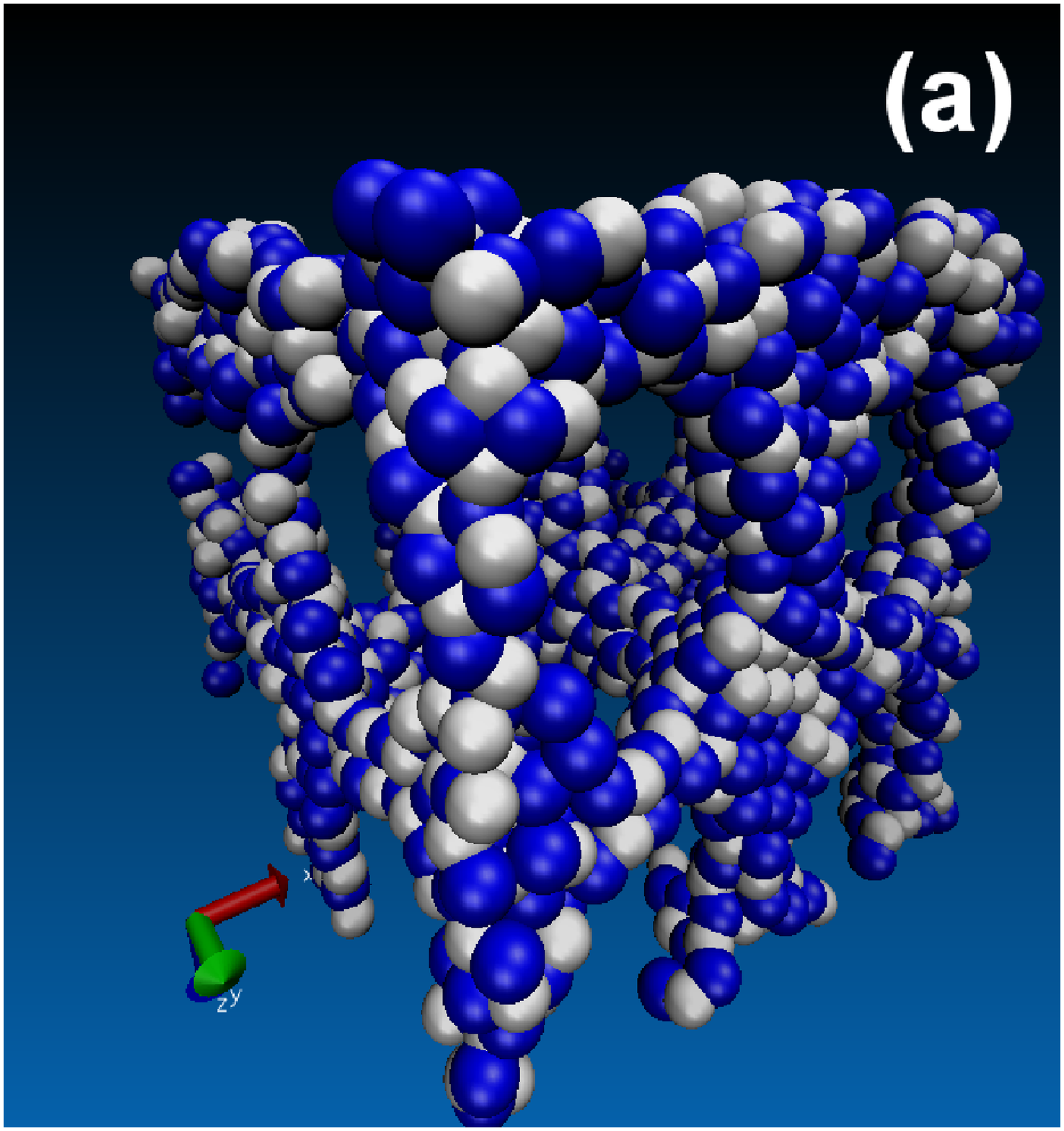} &
\includegraphics[width=1.6in]{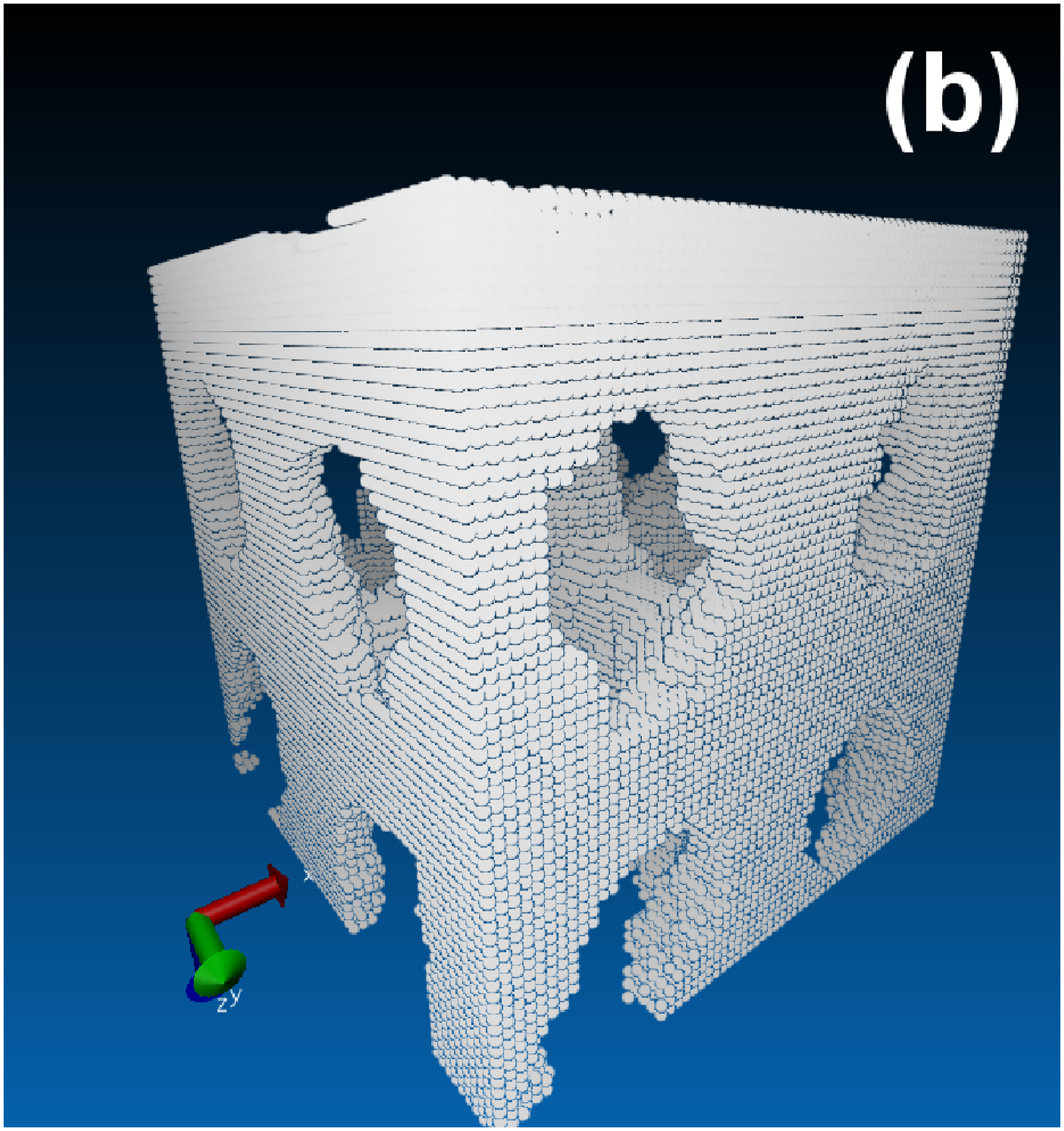}
\end{array}$
\end{center}
\caption{(Color online) Sample transformation of a nuclear structure to a
corresponding polyhedron.  The structure corresponds to a case
with $x=0.5$, $\rho=0.33 \ fm^{-3}$ and $T=0.1 \ MeV$.} \label{Euler}
\end{figure}

In general, $V,\ A, \ H$ plus an interesting construct known as
the Euler characteristic are collectively known as the ``Minkowsky
functionals''; according to integral geometry the morphological
properties of $3D$ objects can be completely described in terms of
them.

The XVIII century work of Euler-L'Huilier showed that, independent
of the shape of any polyhedra, when deducting the number of edges
from the number of vertices and adding the number of faces it
always yields $2$ plus twice the number of cavities, quantity now
known as the ``Euler characteristic'', $\chi$. Although this
previous property is for solids bounded by plane surfaces, it also
applies in any $3D$ surfaces with $\chi$ related to the total
curvature of the surface through the Gauss-Bonnet theorem.

In topological terms, two orientable closed surfaces are
homeomorphic to each other if their Euler characteristics are the
same. Conversely, two homomeorphic closed surfaces will always
have the same value of $\chi$. Therefore, since our pasta niblets
are all orientable, their Euler characteristic completely
classifies them up to an homeomorphism; adding the rest of the
Minkowsky functionals eliminates such redundancy and guarantees a
complete classification of the pasta shapes.

\begin{figure}[t]  
\begin{center}$
\begin{array}{cc}
\includegraphics[width=1.6in]{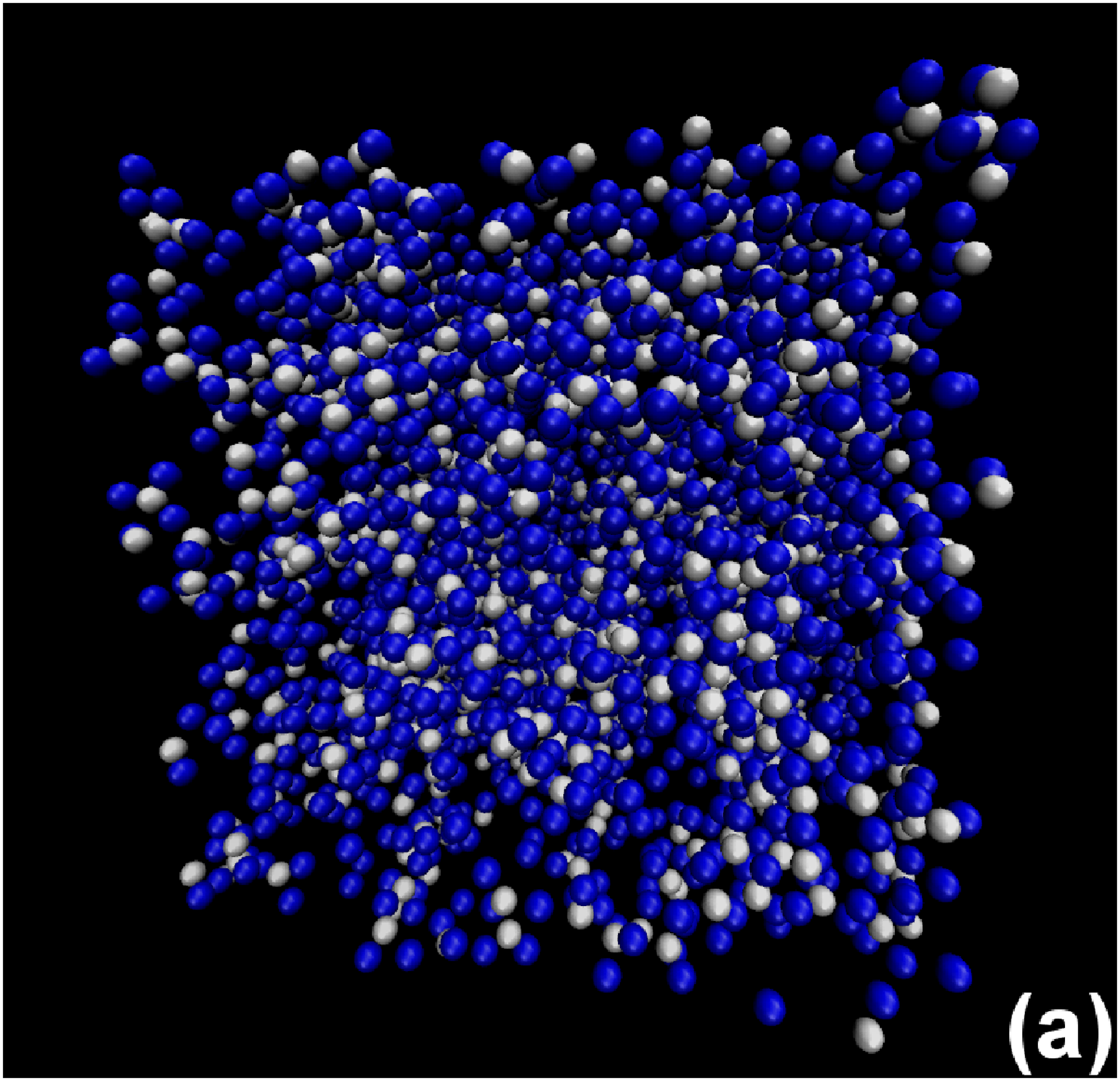} &
\includegraphics[width=1.6in]{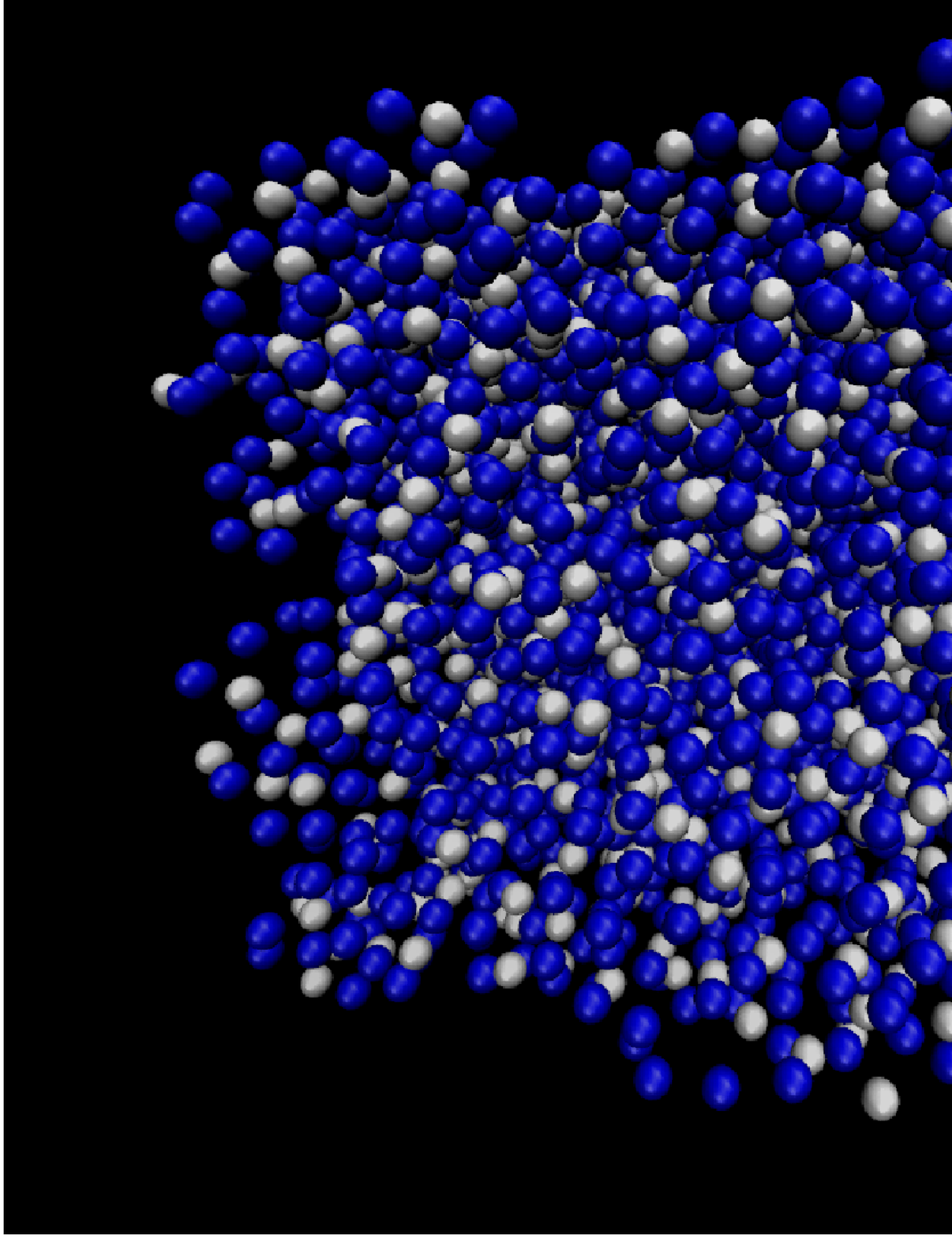}
\end{array}$
\end{center}
\caption{(Color online) Spatial configurations formed under $T=0.4$ and $\rho=0.045 \ fm^{-3}$ (left) and $T=1.0$ and $\rho=0.072 \ fm^{-3}$ (right),
both for $x=0.3$.} \label{configs}
\end{figure}

An immediate problem is the fact that the nuclear clusters are not
polyhedra and do not even form closed surfaces.  This obstacle,
however, can be circumvented by replacing the nuclear structure
with a scaffold-like armature composed of cells enclosing each one
nucleon.  In our case this is done through the algorithm of
Michielsen and De Raedt~\cite{michielsen} which has already been
used in the study of stellar crusts albeit in a different
methodology~\cite{gw-2002}.

Synoptically, the simulation volume is subdivided into a mall of
cubic cells. Those cells which contain the coordinates of a
nucleon are kept while the rest are deleted. The sizes of the
cells are made smaller than the nearest neighbor distance found in
$g(r)$ to enforce a one-particle-per-cell occupation, but not too
small as to avoid creating spurious cavities between neighboring
nucleons. It is on this imaginary platform that the Minkowsky
functionals of the nuclear structure are computed.

\begin{figure}[t]  
\begin{center}$
\begin{array}{cc}
\includegraphics[width=2.6in]{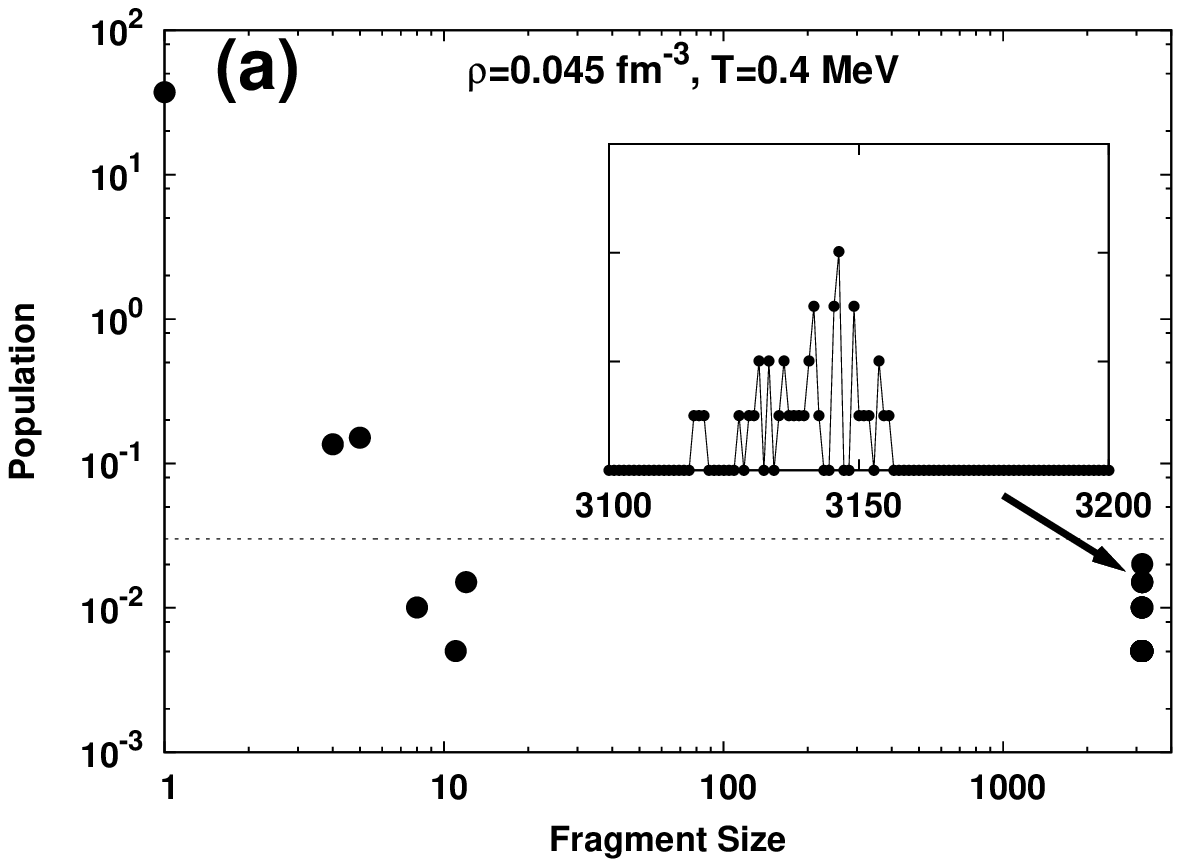} \\
\includegraphics[width=2.6in]{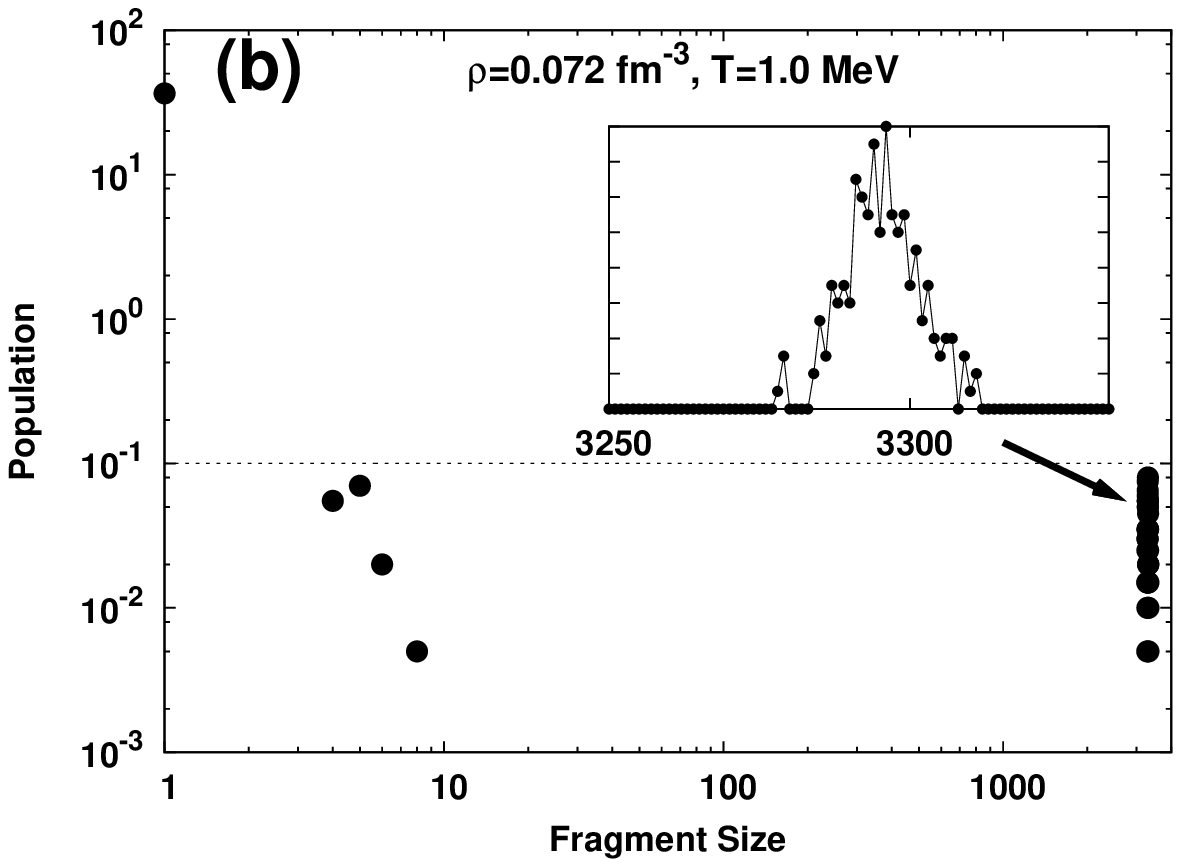}
\end{array}$
\end{center}
\caption{Multiplicites corresponding to figure~\ref{configs},
although not identical they are very similar.} \label{configs2}
\end{figure}

In general, $\chi$ equals the number of regions of connected grid
cells minus the number of completely enclosed regions of empty
grid cells. Two grid cells are connected if they are immediate
neighbors, next-nearest neighbors, or are connected by a chain of
occupied grid cells. Characterizing the connected structure by its
number of occupied cubes, $n_c$, edges, $n_e$, faces, $n_f$, and
vertices, $n_v$, including possible contributions from the
interior of the structure, the Minkowski functionals can be
calculated through~\cite{michielsen}
\begin{eqnarray}
V = n_c, \ && A = -6n_c+2n_f, \nonumber \\ 2B = 3n_c-2n_f+n_c, \
&& \chi=-n_c+n_f-n_e+n_v \nonumber
\end{eqnarray}
where $V$ stands for the volume, $A$ for the area, $B$ for the
mean breadth $B$ and $\chi$ for the Euler number; the mean breadth
measures of the average ``size'' of a body and it is related to
the integral mean curvature $H$ mentioned before.
Figure~\ref{Euler} shows a typical nuclear structure along with
the grid constructed around it; the values of the Minkowski
functionals obtained from such grid are Curvature $=215$ and Euler
$=-17$, as we will see next such shape can be classified as a
``jungle gym''.

\begin{figure}[t]  
\begin{center}$
\begin{array}{cc}
\includegraphics[width=3.4in]{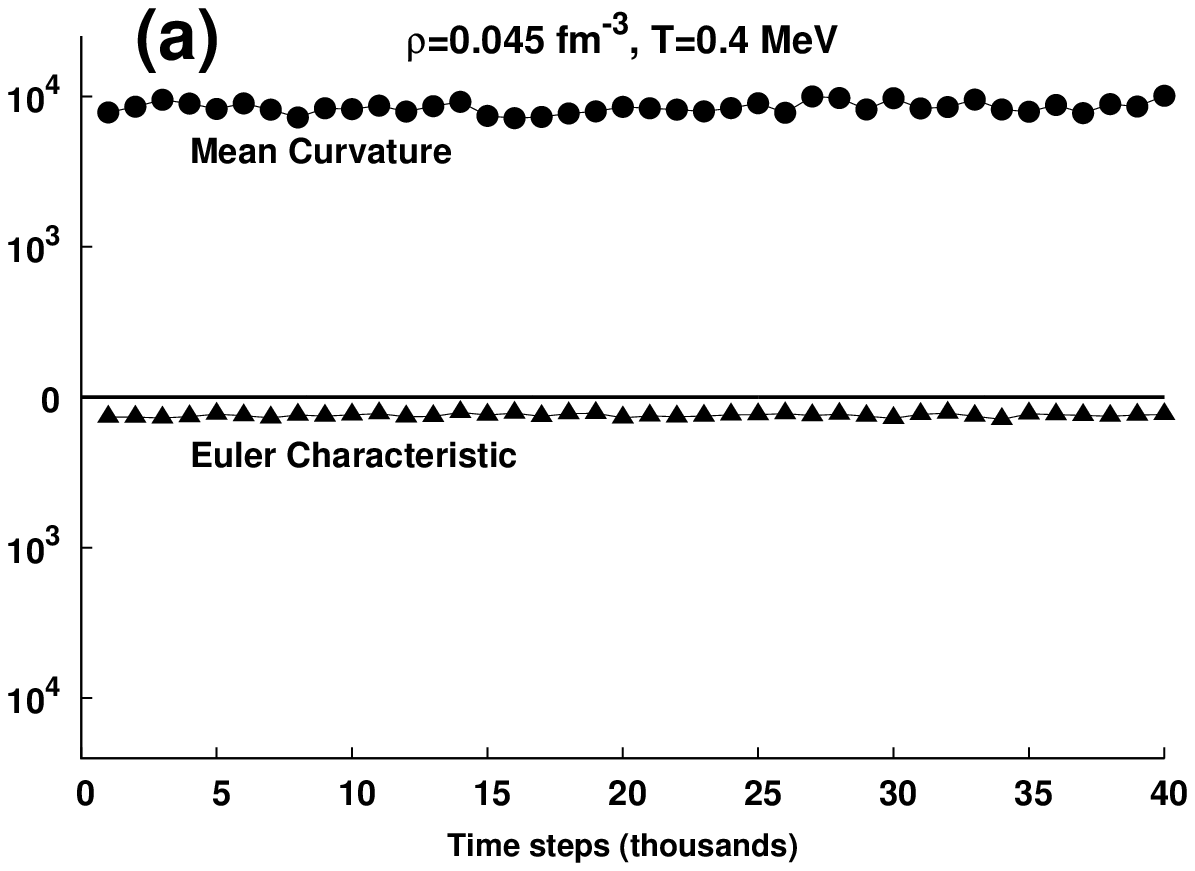} \\
\includegraphics[width=3.4in]{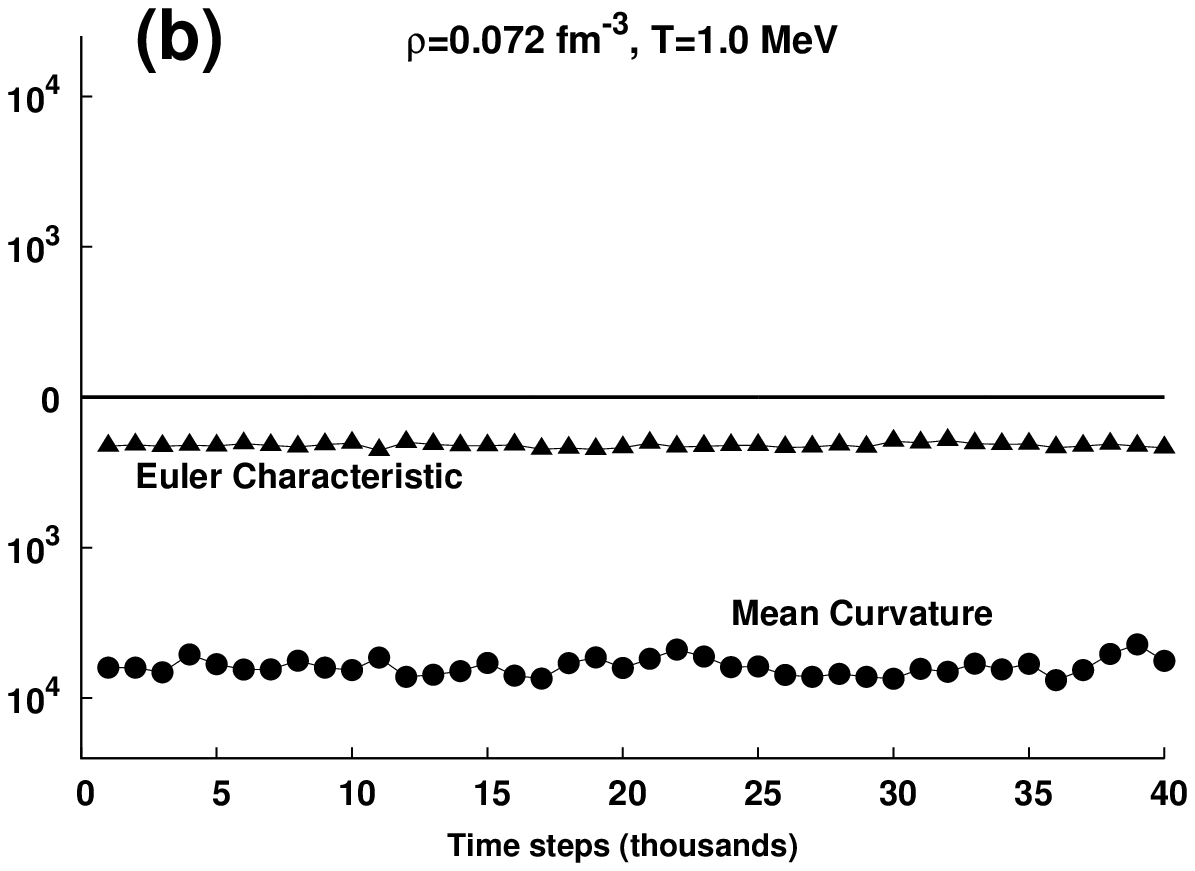}
\end{array}$
\end{center}
\caption{Euler and curvature of the structures shown in
figure~\ref{configs}, the difference between the two cases is easy
to spot.} \label{configs3}
\end{figure}

\subsection{Topological classification of the pasta}

To illustrate the use of topology to classify the pasta shapes let us use two seemingly similar structures obtained with $x=0.3$ but at different densities and temperatures, namely $\rho=0.045 \ fm^{-3}$ and $T=0.4 \ MeV$
and $\rho=0.072 \ fm^{-3}$ $T=1.0 \ MeV$.  The spatial
configurations of these two cases, practically identical to the eye,
are presented in figure~\ref{configs}.  Although there are minor
differences (see insets), figure~\ref{configs2} shows that both
configurations have very similar mass multiplicities.

\begin{figure}[h]  
\begin{center}$
\begin{array}{cc}
\includegraphics[width=1.5in]{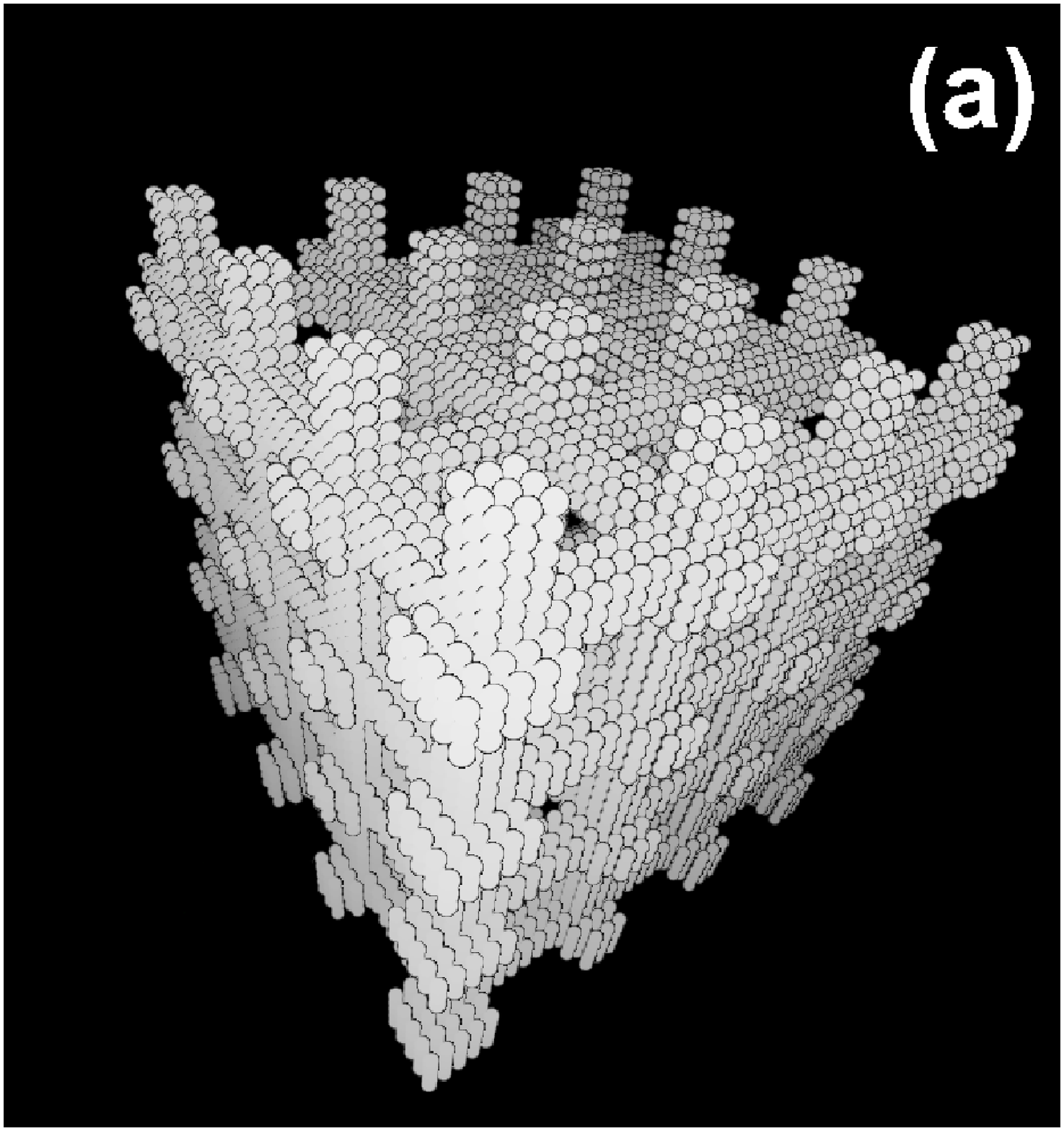} &
\includegraphics[width=1.5in]{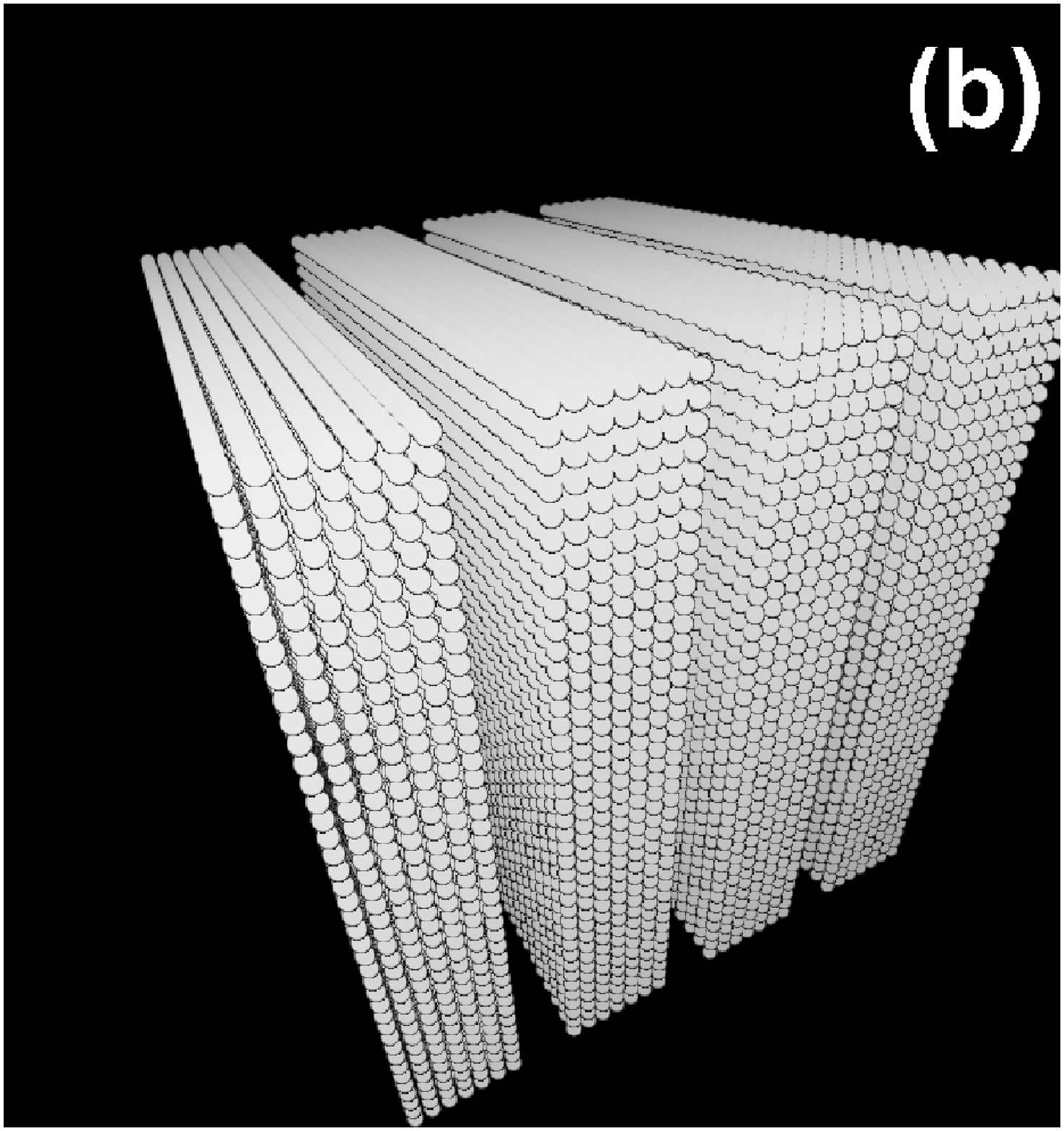} \\ \includegraphics[width=1.5in]{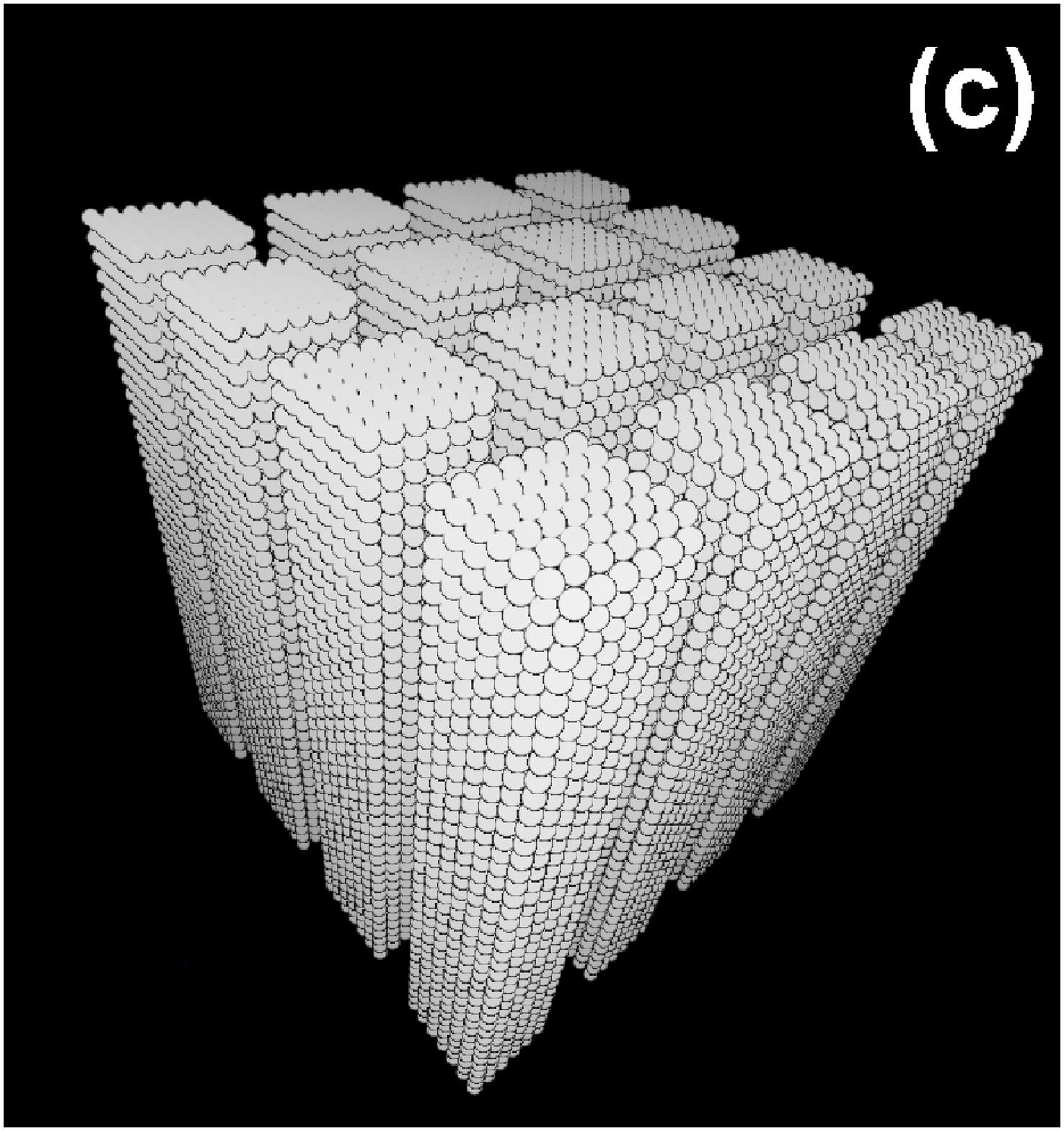} &
\includegraphics[width=1.5in]{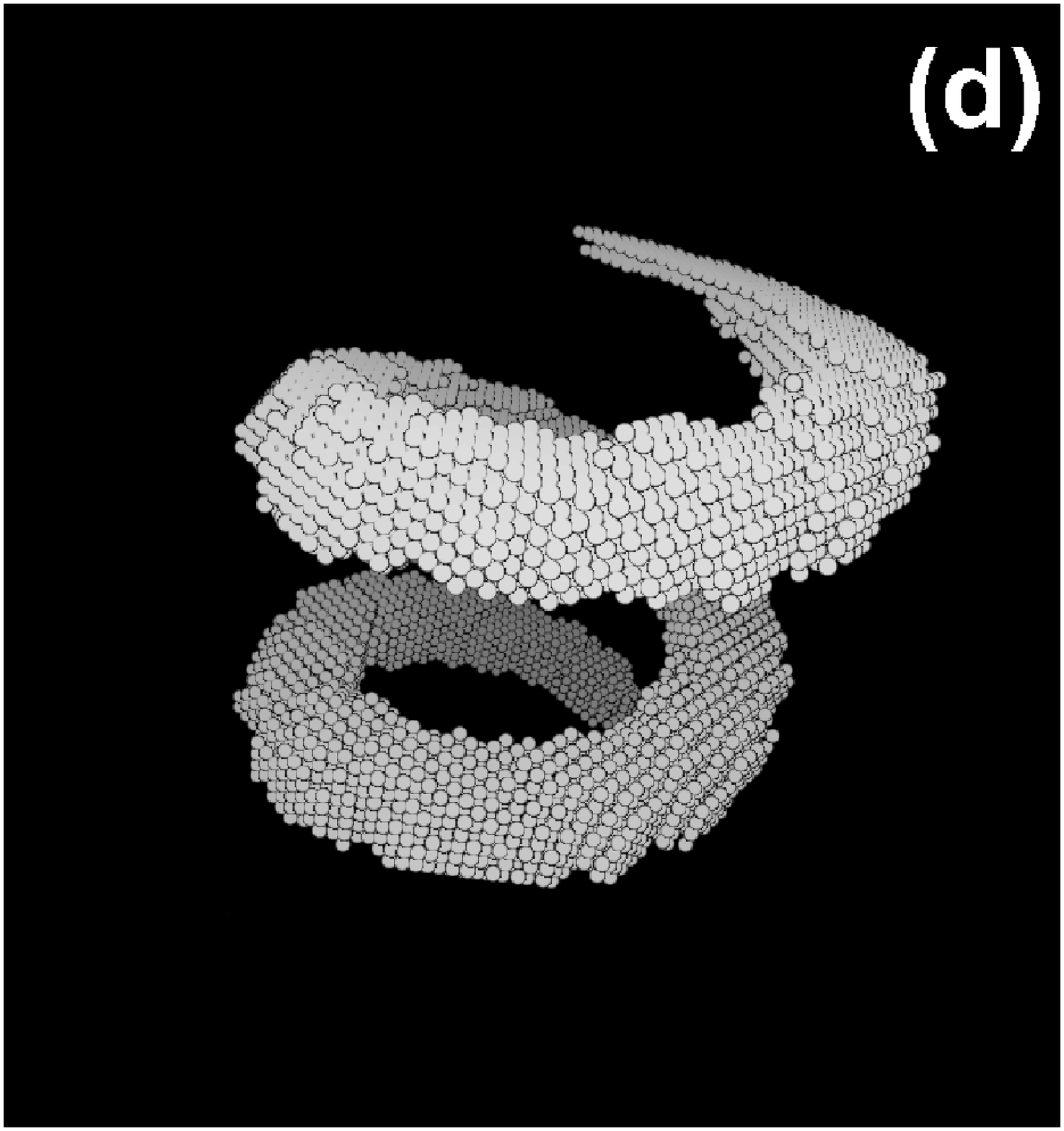}
\end{array}$
\end{center}
\caption{(Color online) Typical artificial ``pastas'' used to test the classification
powers of the Euler-Curvature combination.  Respectively they are
the ``jungle gym'' (top left), ``lasagnas'' (top right),
``straight spaghettis'' (bottom left) and a
``curled spaghetti''.}\label{Artificial}
\end{figure}

The difference between the configurations, however, surfaces when
we calculate the corresponding curvature and Euler numbers.
Figure~\ref{configs3} shows the evolution of the curvature and of
the Euler number as the simulations advances after equilibration.
Clearly shown are the inverted values of Euler and the curvature
for the two cases; while the curvature is positive (with less
cavities than bellies) and large in the low $T$ low $\rho$ case,
it becomes negative (with more internal cavities than tummies) and
smaller in the opposite case.

\begin{figure}[h]  
\begin{center}
\includegraphics[width=3.5in]{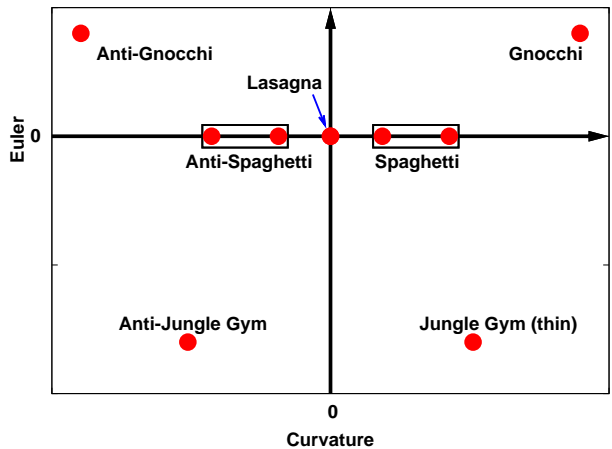}
\end{center}
\caption{(Color online) Systematic classification of the artificial structures in
terms of the curvature and Euler number.} \label{ECPlane}
\end{figure}

To investigate this point further, we created artificial
structures in the form of gnocchi, spaghetti, lasagna and
crossed-lasagnas, which we call ``jungle gym'', and their inverse
structures (with voids replacing particles and viceversa) and
calculated the values of the two topological variables; some of
the structures used are shown in figure~\ref{Artificial} and their
location in the Curvature-Euler plane in figure~\ref{ECPlane}. The
magnitudes shown are determined by the size of the structure as
well as the digitization cell size. In general one can conclude
that lasagnas tend to lie near the origin, spaghettis have near
zero Euler numbers and positive curvatures, gnocchis have positive
curvatures and Euler numbers, and ``jungle gyms'' positive Euler
number and negative curvature; all anti-structures reverse the
curvature but maintain the Euler characteristic.  All cases
calculated at all $x$ values, densities and temperatures were
observed to satisfy this classification.

\begin{table}[ht]
\centering  
\caption{Classification Curvature - Euler}
\begin{tabular}{c | r c c | r c c} 
\hline \hline                        
Density & \multicolumn{3}{c|}{$x = 0.5$} & \multicolumn{3}{c}{$x = 0.3$} \\
[0.5ex]
($fm^{-3}$) & Curvature & Euler & Topology & Curvature & Euler & Topology \\
[0.5ex]
\hline                  
0.01 & (a) 100 & 100 & G & (A) 96 & 27 & G \\ 
0.015 & 73 & 50 & G  & 92 & 7 & G-S \\
0.018 & 58 & 17  & G-S & 79 & -9 & S\\
0.021 & 36 & -25 & S-J & & & \\
0.024 & 22 & -28 & S-J & 58 & -18 & J-S  \\
0.026 &    &     &  & 51 & -39 &  J \\
0.027 &  9 & -42 & J-L & 47 & -37 & J  \\
0.03  & 10 & -39 & J-L & 48 & -7 &  S \\
0.033 &  9 & -47 & J & 18 & -75 & J  \\
0.036 &  8 & -42 & J & & &   \\
0.039 & -11 & -6 & L-AJ & & &   \\
0.042 & -15 & -8 & L-AJ & & &   \\
0.045 &  1 & -33 & L-J & -54 & -100 & AJ  \\
0.048 & -5 & -11 & L & & &   \\
0.051 & -7 & -17 & AS-AJ & -94 & -41 & AJ  \\
0.054 & -1 & -11 & L-AJ & & &   \\
0.057 & -9 & -30 & AJ & & &   \\
0.06 & -9 & -17 & AJ & -100 & 66 & AG  \\
0.063 & -10 & -30 & AJ & & &   \\
0.072 & -12 & -8 & AS-AJ & (L) -60 & 90 & AG  \\
0.084 & (t) -19 & -8 & AJ & & &   \\ [1ex]      
\hline 
\end{tabular}
\label{table1} 
\end{table}

For instance, the structure in figure~\ref{Euler} with curvature
$215$ and Euler number $-17$ is clearly a ``jungle gym''; it must
be remarked that --to our knowledge-- this is the first time this
type of structure has been reported. Likewise, the structure in
the left panel of figure~\ref{configs}, with both negative
curvature and Euler number, can be classified as an ``anti-jungle
gym'', whereas the accompanying structure with positive curvature
and negative Euler number would have to be classified as a
``jungle gym''.

\begin{figure}[h]  
\begin{center}
\includegraphics[width=3.55in]{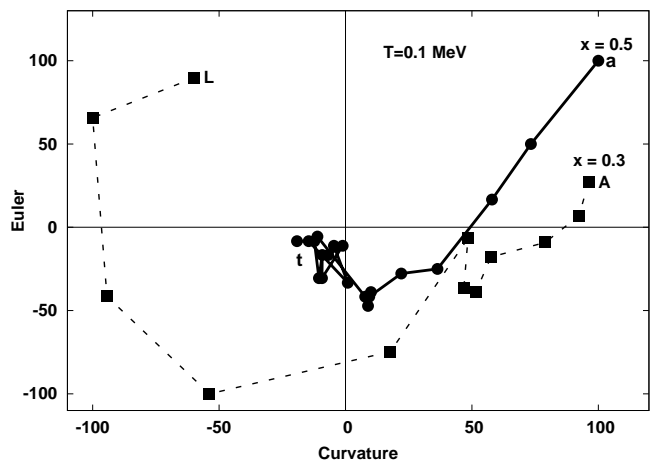}
\end{center}
\caption{Curvature-Euler coordinates of the structures listed in
table~\ref{table1}. Labels ``A'', ``a'', ``L'' and ``t''
correspond, respectively, to the initial and final points in the
table.} \label{curv-euler}
\end{figure}

Table~\ref{table1} shows the classification of several of the
structures obtained in our study at $T=0.1 \ MeV$ and at the
listed densities, and figure~\ref{curv-euler} shows their location
in the $C-E$ plane; the $x=0.5$ column of the table and the
circular dots on the figure correspond to the structures presented
in figure~\ref{Pasta}. As the absolute magnitude of the curvature
and Euler number depends on the overall size of the structure,
i.e. on the number of particles used, the data in
table~\ref{table1} were normalized to have maximum absolute values
of $100$. In the table, the classifications are abbreviated as G
for gnocchi, J for jungle gym, L for lasagna, S for spaghetti, and
AG, AJ, AL and AS for the reverse structures.

\section{Concluding remarks}
And thus we have reached our objective.  The combination of
curvature and Euler number and a proper recognition of fragments appears to be robust enough as to uniquely classify
the shapes attained by the nucleons at densities, temperatures and
isospin content of interest in the study of neutron star crusts.
In obtaining this result, the classical molecular dynamics and
associated tools (cluster recognition algorithms, persistence,
etc.) proved to be a valuable method, which we now plan to
exploit.

In future investigations we will apply this method to an in-depth
study of the origin of clustering.  As hinted in~\cite{Dor11}, the
role of the long range Coulomb interaction has not been fully explored;
current exploratory runs are indicating that Coulomb merely shifts
the space scales a bit but is not an ``if-and-only-if''
requirement for the formation of clusters. Along the same lines,
as the present $CMD$ model can function both with a stiff and
medium compressibility potentials, we also plan to investigate the
role of the equations of state in the formation of the pasta
structures.

\begin{acknowledgments}
C.O.D. is a member of the ``Carrera del Investigador'' CONICET
supported by CONICET through grant PIP5969, and acknowledges the warm hospitality of the University of Texas at El Paso.  J.A.L. Acknowledges support from grant NSF-PHY 1066031, thanks Dr. Jorge Piekarewicz for suggesting the use of $CMD$ to study the nuclear pasta, Dr. L.G. Valdez S\'anchez for clarifying observations about our use of topological objects, and the hospitality of the University of Buenos Aires and of the Lawrence Berkeley Laboratory where this manuscript was completed.

\end{acknowledgments}

\end{document}